\documentclass{llncs}

\usepackage[svgnames, x11names]{xcolor}

\usepackage[normalem]{ulem} 

\usepackage{listings}

\lstdefinelanguage{XML}
{
basicstyle=\ttfamily\footnotesize,
  morestring=[b]",
  moredelim=[s][\bfseries\color{Maroon}]{<}{\ },
  moredelim=[s][\bfseries\color{Maroon}]{</}{>},
  moredelim=[l][\bfseries\color{Maroon}]{/>},
  moredelim=[l][\bfseries\color{Maroon}]{>},
  morecomment=[s]{<?}{?>},
  morecomment=[s]{<!--}{-->},
  commentstyle=\color{gray},
  stringstyle=\color{blue},
  identifierstyle=\color{red}
}
\usepackage[pdftex]{graphicx}
\graphicspath{{./figures/}}
\DeclareGraphicsExtensions{.pdf}

\usepackage[cmex10]{amsmath}
\usepackage{amssymb}
\usepackage{mathtools}

\usepackage[caption=false,font=scriptsize]{subfig}

\usepackage{algorithmicx}
\usepackage{algpseudocode}
\usepackage[ruled]{algorithm}
\definecolor{light-gray}{gray}{0.75}
\algrenewcommand{\algorithmiccomment}[1]{\hskip3em{{\footnotesize \textcolor{light-gray}{$\blacktriangleright$}}} #1}

\usepackage{multirow}

\usepackage{xspace}
\usepackage[nocompress]{cite}

\usepackage{enumitem}

\begin{document}
\mainmatter              
\title{Benchmarking Distributed Stream Processing Platforms for IoT Applications}
\titlerunning{IoT Benchmark for Stream Processing}  
%
\author{Anshu Shukla and Yogesh Simmhan} 
\authorrunning{Shukla, et al.} 
%
%

\institute{Indian Institute of Science, Bangalore India\\
\email{shukla@grads.cds.iisc.ac.in, simmhan@cds.iisc.ac.in}\\ 
}

\maketitle              

\begin{abstract}
Internet of Things (IoT) is a technology paradigm where millions of sensors monitor, and help inform or manage, physical, environmental and human systems in real-time. The inherent closed-loop responsiveness and decision making of IoT applications makes them ideal candidates for using low latency and scalable stream processing platforms. Distributed Stream Processing Systems (DSPS) are becoming essential components of any IoT stack, but the efficacy and performance of contemporary DSPS have not been rigorously studied for IoT data streams and applications. Here, we develop a benchmark suite and performance metrics to evaluate DSPS for streaming IoT applications. The benchmark includes $13$ common IoT tasks classified across various functional categories and forming micro-benchmarks, and two IoT applications for statistical summarization and predictive analytics that leverage various dataflow compositional features of DSPS. These are coupled with stream workloads sourced from real IoT observations from smart cities. We validate the IoT benchmark for the popular Apache Storm DSPS, and present empirical results.
\end{abstract}

\section{Introduction}
\label{sec:intro}
Internet of Things (IoT) refers to a technology paradigm wherein ubiquitous sensors numbering in the billions will able to monitor physical infrastructure and environment, human beings and virtual entities in real-time, process both real-time and historic observations, and take actions that improve the efficiency and reliability of systems, or the comfort and lifestyle of society. The technology building blocks for IoT have been ramping up over a decade, with research into pervasive and ubiquitous computing~\cite{zaslavsky2013internet}, and sensor networks~\cite{chandrasekaran:cidr:2003} forming precursors. Recent growth in the capabilities of high-speed mobile (e.g., 2G/3G/4G) and \emph{ad hoc} (e.g., Bluetooth) networks~\cite{ericsson-report}, smart phones, affordable sensing and crowd-sourced data collection~\cite{data:city}, Cloud data-centers and Big Data analytics platforms
 have all contributed to the current inflection point for IoT. 

Currently, the IoT applications are often manifest in vertical domains, such as demand-response optimization and outage management in \emph{smart grids}~\cite{smart-grid-dr-iot}, or fitness and sleep tracking and recommendations by \emph{smart watches and health bands}~\cite{fitness-iot}. 
The IoT stack for such domains is tightly integrated to serve specific needs, but typically operates on a closed-loop \emph{Observe Orient Decide Act (OODA)} cycle, where sensors communicate time-series observations of the (physical or human) system to a central server or the Cloud for analysis, and the analytics drive recommendations that are enacted on, or notified to, the system to improve it, which is again observed and so on. In fact, this \emph{closed-loop} responsiveness is one of the essential design characteristics of IoT applications.


This low-latency cycle makes it necessary to process data streaming from sensors at fine spatial and temporal scales, in \emph{real-time}, to derive actionable intelligence. In particular, this streaming analytics has be to done at massive scales (millions of sensors, thousands of events per second) from across distributed sensors, requiring large computational resources. \emph{Cloud computing} offers a natural platform for scalable processing of the observations at globally distributed data centers, and sending a feedback response to the IoT system at the edge. 

Recent \emph{Big Data platforms} like Apache Storm~\cite{toshniwal:sigmod:2014} and Spark~\cite{zaharia:usenix:2012} provide an intuitive programming model for composing such streaming applications, with a scalable, low-latency execution engine designed for commodity clusters and Clouds. 
These \emph{Distributed Stream Processing Systems (DSPS)} are becoming essential components of any IoT stack to support online analytics and decision-making for IoT applications. In fact, reference IoT solutions from Cloud providers like Amazon AWS\footnote{https://aws.amazon.com/iot/how-it-works/} and Microsoft Azure\footnote{https://www.microsoft.com/en-in/server-cloud/internet-of-things/overview.aspx} include their proprietary stream and event processing engines as part of the IoT analytics architecture.

Shared-memory stream processing systems~\cite{chandrasekaran:cidr:2003, aurora} have been investigated over a decade back for wireless sensor networks, with community benchmarks such as \emph{Linear Road}~\cite{arasu:vldb:2004} being proposed. But there has not been a detailed review of, or benchmarks for, \emph{distributed} stream processing for IoT domains. In particular, the efficacy and performance of contemporary DSPS, which were originally designed for social network and web traffic~\cite{toshniwal:sigmod:2014}, have not been rigorously studied for \emph{IoT data streams and applications}. We address this gap in this paper.

We develop a benchmark suite for DSPS to evaluate their effectiveness for streaming IoT applications. The proposed workload is based on common building-block tasks observed in various IoT domains for real-time decision making, and the input streams are sourced from real IoT observations from smart cities.

Specifically, we make the following contributions:
\begin{enumerate}
\item We classify different characteristics of streaming applications and their data sources, in \S~\ref{sec:features}. We propose categories of tasks that are essential for IoT applications and the key features that are present in their input data streams.
\item We identify performance metrics of DSPS that are necessary to meet the latency and scalability needs of streaming IoT applications, in \S~\ref{sec:metrics}.
\item We propose an IoT Benchmark for DSPS based on representative \emph{micro-benchmark tasks}, drawn from the above categories, in \S~\ref{sec:benchmark}. Further, we design two reference IoT applications -- for \emph{statistical analytics} and \emph{predictive analytics} -- composed from these tasks. We also offer real-world streams with different distributions on which to evaluate them.
\item We run the benchmark for the popular Apache Storm DSPS, and present empirical results for the same in \S~\ref{sec:results}.
\end{enumerate}

Our contributions here will allow IoT applications to evaluate if current and future DSPS meet their performance and scalability needs, and offer a baseline for Big Data researchers and developers to uniformly compare DSPS platforms for different IoT domains.

\section{Background and Related Work}
\label{sec:related}

Stream processing systems allow users to compose applications as a dataflow graph, with task vertices having some user-defined logic, and streaming edges passing messages between the tasks, and run these applications continuously over incoming data streams. 
%
Early data stream management systems (DSMS) were motivated by sensor network applications, that have similarities to IoT~\cite{carney:vldb:2002,chen:sigmod:2000}. They supported continuous query languages with operators such as join, aggregators similar to SQL, but with a temporal dimension using windowed-join operations. These have been extended to distributed implementations~\cite{balazinska:tods:2008} and complex event processing (CEP) engines for detecting sequences and patterns.  

Current distributed stream processing systems (DSPS) like Storm and Spark Streaming~\cite{toshniwal:sigmod:2014,zaharia:usenix:2012,neumeyer:icdmw:2010} leverage Big Data fundamentals, running on commodity clusters and Clouds, offering weak scaling, ensuring robustness, and supporting fast data processing over thousands of events per second. They do not support native query operators and instead allow users to plug in their own logic composed as dataflow graphs executed across a cluster. While developed for web and social network applications, such fast data platforms have found use in financial markets, astronomy, and particle physics. IoT is one of the more recent domains to consider them. 

Work on DSMS spawned the Linear Road Benchmark (LRB)~\cite{arasu:vldb:2004} that was proposed as an application benchmark. In the scenario, DSMS had to evaluate toll and traffic queries over event streams from a virtual toll collection and traffic monitoring system. This parallels with current smart transportation scenarios. However, there have been few studies or community efforts on benchmarking DSPS, other than individual evaluation of research prototypes against popular DSPS like Storm or Spark.

These efforts define their own measures of success -- typically limited to throughput and latency -- and use generic workloads such as enron email dataset  with no-operation as micro-benchmark to compare InfoSphere Streams ~\cite{nabi:streams:2014} and Storm.SparkBench~\cite{agrawal:sparkbench:2015} uses two streaming applications, Twitter popular tag retrieving data from the twitter website to calculate most popular tag every minute and PageView over synthetic user clicks  to get various statistics using spark.



Stream Bench~\cite{lu:ucc:2014} has proposed 7 micro-benchmarks on 4 different synthetic workload suites generated from real-time web logs and network traffic to evaluate DSPS. Metrics including performance, durability and fault tolerance are proposed. The benchmark covers different dataflow composition patterns and common tasks like grep and wordcount. While useful as a generic streaming benchmark, it does not consider aspects unique to IoT applications and streams.

SparkBench~\cite{agrawal:sparkbench:2015} is a framework-specific benchmark for Apache Spark, and includes four categories of applications from domains spanning Graph computation and SQL queries, with one on streaming applications supported by Spark Streaming. The benchmark metrics include CPU, memory, disk and network IO, with the goal of identifying tuning parameters to improve Spark's performance. 

CEPBen~\cite{li:tpctc:2014} evaluates the performance of CEP systems based of the functional behavior of queries. It shows the degree of complexity of CEP operations like filter, transform and pattern detection. The evaluation metrics consider event processing latency, but ignore network overheads and CPU utilization. Further, CEP applications rely on a declarative query syntax to match event patterns rather than a dataflow composition based on user-logic provided by DSPS.

In contrast, the goal for this paper is to develop relevant micro- and application-level benchmarks for evaluating DSPS, specifically for \emph{IoT workloads} for which such platforms are increasingly being used. Our benchmark is designed to be \emph{platform-agnostic}, \emph{simple} to implement and execute within diverse DSPS, and \emph{representative} of both the application logic and data streams observed in IoT domains. This allows for the performance of DSPS to be independently and reproducibly verified for IoT applications. 

There has been a slew of Big Data benchmarks that have been developed recently in the context of processing high volume (i.e., MapReduce-style) and enterprise/web data that complement our work. 
\emph{Hibench}~\cite{huang:hibench:2010} is a workload suite for evaluating Hadoop with popular micro-benchmarks like Sort, WordCount and TeraSort, MapReduce applications like Nutch Indexing and PageRank, and machine learning algorithms like K-means Clustering. \emph{BigDataBench}~\cite{gao:bigdatabench:2013} analyzes workloads from social network and search engines, and analytics algorithms like Support Vector Machine (SVM) over structured, semi-structured and unstructured web data. Both these benchmarks are general purpose workloads that do not target any specific  domain, but MapReduce platforms at large. 

\emph{BigBench}~\cite{ghazal:acm:2013} uses a synthetic data generator to simulate enterprise data found in online retail businesses. It combines structured data generation from the TPC-DS benchmark~\cite{nambiar:vldb:2006}, semi-structured data on user clicks, and unstructured data from online product reviews. Queries cover data \emph{velocity} by processing periodic refreshes that feed into the data store, \emph{variety} by including free-text user reviews, and \emph{volume} by querying over a large web log of clicks. We take a similar approach for benchmarking fast data platforms, targeting the IoT domain specifically and using real public data streams. 






There has been some recent work on benchmarking IoT applications. In particular, the generating large volumes of synthetic sensor data with realistic values is challenging, yet required for benchmarking. \emph{IoTAbench}~\cite{arlitt:icpe:2015} provides a scalable synthetic generator of time-series datasets. It uses a Markov chain model for scaling the time series with a limited number of inputs such that important statistical properties of the stream is retained in the generated data. They have demonstrated this for smart meter data. The benchmark also includes six SQL queries to evaluate the performance of different query platforms on the generated dataset. Their emphasis is more on the data characteristics and content, which supplements our focus on the systems aspects of the executing platform. 

CityBench~\cite{ali:citybench:2015} is a benchmark to evaluate RDF stream processing systems. They include different generation patterns for smart city data, such as traffic vehicles, parking, weather, pollution, cultural and library events, with changing event rates and playback speeds. They propose fixed set of semantic queries over this dataset, with concurrent execution of queries and sensor streams. Here, the target platform is different (RDF database), but in a spirit as our work. 






\section{Characteristics of Streaming IoT Applications}
\label{sec:features}

In this section, we review the common application composition capabilities of DSPS, and the dimensions of the streaming applications that affect their performance on DSPS. These semantics help define and describe streaming IoT applications based on DSPS capabilities. Subsequently in this section, we also categorize IoT tasks, applications and data streams based on the domain requirements. Together, these offer a search space for defining workloads that meaningfully and comprehensively validate IoT applications on DSPS. 



\subsection{Dataflow Composition Semantics}
\label{subsec:dataflow}
DSPS applications are commonly composed as a \emph{dataflow graph}, where vertices are user provided \emph{tasks} and directed edges are refer to \emph{streams of messages} that can pass between them. The graph need not be acyclic. Tasks in the dataflows can execute zero or more times, and a task execution usually depends on data-dependency semantics, i.e, when ``adequate'' inputs are available, the task executes. However, there are also more nuanced patterns that are supported by DSPS that we discuss.
\emph{Messages} (or events or tuples) from/to the stream are consumed/produced by the tasks. 
DSPS typically treat the messages as opaque content, and only the user logic may interpret the message content. However, DSPS may assign identifiers to messages for fault-tolerance and delivery guarantees, and some message attributes may be explicitly exposed as part of the application composition for the DSPS to route messages to downstream tasks. 

\emph{Selectivity ratio}, also called \emph{gain}, is the number of output messages emitted by a task on consuming a unit input message, expressed as $\sigma$=\emph{input rate}:\emph{output rate}. Based on this, one can assess whether a task amplifies or attenuates the incoming message rate. It is important to consider this while designing benchmarks as it can have a multiplicative impact on downstream tasks. 




\begin{figure*}[t]
  \centering
	\includegraphics[width=\textwidth]{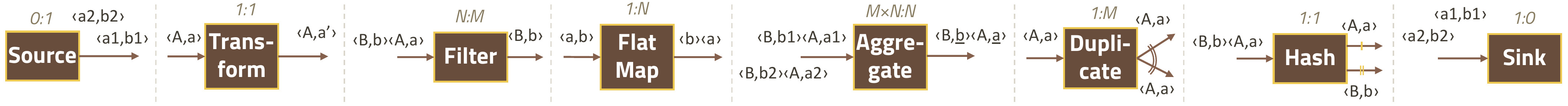}
	\caption{Common task patterns and semantics in streaming applications.}
    \label{fig:semantics}
\end{figure*}

There are message generation, consumption and routing semantics associated with tasks and their dataflow composition. Fig.~\ref{fig:semantics} captures the basic \emph{composition patterns} supported by modern DSPS. \texttt{Source} tasks have only outgoing edge(s), and these tasks encapsulate user logic to generate or receive the input messages that are passed to the dataflow. Likewise, \texttt{Sink} tasks have only incoming edge(s) and these tasks react to the output messages from the application, say, by storing it or sending an external notification. 

\texttt{Transform} tasks, sometimes called \emph{Map} tasks, generate one output message for every input message received ($\sigma=1:1$). Their user logic performs a transformation on the message, such as changing the units or projecting only a subset of attribute values. \texttt{Filter} tasks allow only a subset of messages that they receive to pass through, optionally performing a transformation on them ($\sigma=N:M$, $N \ge M$). Conversely, a \texttt{FlatMap} consumes one message and emits multiple messages ($\sigma=1:N$). An \texttt{Aggregate} pattern consumes a \emph{window} of messages, with the window width provided as a \emph{count} or a \emph{time} duration, and generates one or more messages that is an aggregation over each message window ($\sigma=N:1$).

When a task has multiple outgoing edges, routing semantics on the dataflow control if an output message is \emph{duplicated} onto all the edges, or just one downstream task is selected for delivery, either based on a \emph{round robin} behavior or using a \emph{hash function} on an attribute in the outgoing message to decide the target task. 
Similarly, multiple incoming streams arriving at a task may be \emph{merged} into a single interleaved message stream for the task. Or alternatively, the messages coming on each incoming stream may be conjugated, based on order of arrival or an attribute exposed in each message, to form a \emph{joined} stream of messages. 

There are additional dimensions of the streaming dataflow that can determine its performance on a DSPS. 
%
%
Tasks may be \emph{data parallel}, in which case, it may be allocated concurrent resources (threads, cores) to process messages in parallel by different instances the task. This is typically possible for tasks that do not maintain state across multiple messages. The \emph{number of tasks} in the dataflow graph indicates the size of the streaming application. Tasks are mapped to computing resources, and depending of their degree of parallelism and resource usage, it determines the cores/VMs required for executing the application. 
%
The \emph{length of the dataflow} is the latency of the critical (i.e., longest) path through the dataflow graph, if the graph does not have cycles. This gives an estimate of the expected latency for each message and also influences the number of network hops a message on the critical path has to take in the cluster. 

\subsection{Input Data Stream Characteristics}
We list a few characteristics of the input data streams that impact the runtime performance of streaming applications, and help classify IoT message streams. 

The \emph{input throughput} in messages/sec is the cumulative frequency at which messages enter the source tasks of the dataflow. Input throughputs can vary by application domain, and are determined both by the number of streams of messages and their individual rates. This combined with the dataflow selectivity will impact the load on the dataflow and the output throughput. 

\emph{Throughput distribution} captures the variation of input throughput over time. In real-world settings, the input data rate is usually not constant and DSPS need to adapt to this. There may be several common data rate distributions besides a \emph{uniform} one. There may be \emph{bursts} of data coming from a single sensor, or a coordinated set of sensors. A \emph{saw-tooth} behavior may be seen in the ramp-up/-down before/after specific events. 
\emph{Normal} distribution are seen with diurnal (day vs. night) stream sources, with \emph{bi-modal} variations capturing peaks during the morning and evening periods of human activity. 

Lastly, the \emph{message size} provides the average size of each message, in bytes. Often, the messages sizes remain constant for structured messages arriving from specific sensor or observation types, but may vary for free-text input streams or those that interleave messages of different types. This size help assess the communication cost of transferring messages in the dataflow.

\subsection{Categories of IoT Tasks and Applications}
IoT covers a broad swathe of domains, many of which are rapidly developing. So, it is not possible to comprehensively capture all possible IoT application scenarios. However, DSPS have clear value in supporting the real-time processing, analytics, decision making and feedback that is intrinsic to most IoT domains. Here, we attempt to categorize these common processing and analytics tasks that are performed over real-time data streams. 

\textbf{Parse.} Messages are encoded on the wire in a standard text-based or binary representation by the stream sources, and need to be parsed upon arrival at the application. Text formats in particular require string parsing by the tasks, and are also larger in size on the wire. The tasks within the application may themselves retain the incoming format in their streams, or switch to another format or data model, say, by projecting a subset of the fields. Industry-standard formats that are popular for IoT domains include CSV, XML and JSON text formats, and EXI and CBOR binary formats.

\textbf{Filter.} Messages may require to be filtered based on specific attribute values present in them, as part of data quality checks, to route a subset of message types to a part of the dataflow graph, or as part of their application logic. Value and band-pass filters that test an attribute's \emph{numerical value ranges} are common, and are both compact to model and fast to execute. Since IoT event rates may be high, more efficient Bloom filters may also be used to process \emph{discrete values} with low space complexity but with a small fraction of false positives.     

\textbf{Statistical Analytics.} Groups of messages within a sequential time or count window of a stream may require to be aggregated as part of the application. The aggregation function may be \emph{common mathematical operations} like average, count, minimum and maximum. They may also be \emph{higher order statistics} such finding outliers, quartiles, second and third order moments, and counts of distinct elements. Statistical \emph{data cleaning} like linear interpolation or denoising using Kalman filters are common for sensor-based data streams. Some tasks may maintain just local state for the window width (e.g., local average) while others may maintain state across windows (e.g., moving average). When the state size grows, here again approximate aggregation algorithms may be used.     

\textbf{Predictive Analytics.} Predicting future behavior of the system based on past and current messages is an important part of IoT applications. Various statistical and machine-learning algorithms may be employed for predictive analytics over sensor streams. The \emph{predictions} may either use a recent window of messages to estimate the future values over a time or count horizon in future, or train models over streaming messages that are periodically used for predictions over the incoming messages. The \emph{training} itself can be an online task that is part of an application. For e.g., ARIMA and linear regression use statistical methods to predict uni- or multi-variate attribute values, respectively. Classification algorithms like decision trees, neural networks and na\"{i}ve Bayes can be trained to map discrete values to a category, which may lead to specific actions taken on the system. External libraries like Weka or statistical packages like R may be used by such tasks.  

\textbf{Pattern Detection.} Another class of tasks are those that identify patterns of behavior over several events. Unlike window aggregation which operate over static window sizes and perform a function over the values, pattern detection matches user-defined predicates on messages that may not be sequential or even span streams, and returned the matched messages. These are often modeled as \emph{state transition automata} or \emph{query graphs}. Common patterns include contiguous or non-contiguous sequence of messages with specific property on each message (e.g., high-low-high pattern over 3 messages), or a join over two streams based on a common attribute value. Complex Event Processing (CEP) engines~\cite{siddhi} may be embedded within the DSPS task to match these patterns.  

\textbf{Visual Analytics.} Other than automated decision making, IoT applications often generate \emph{charts and animations} for consumption by end-users or system managers. These visual analytics may be performed either at the client, in which case the processed data stream is aggregated and provided to the users. Alternatively, the streaming application may itself periodically generate such plots and visualizations as part of the dataflow, to be hosted on the web or pushed to the client. Charting and visualization libraries like D3.js or JFreeChart may be used for this purpose.  

\textbf{IO Operations.} Lastly, the IoT dataflow may need to access external storage or messaging services to access/push data into/out of the application. These may be to store or load trained models, archive incoming data streams, access historic data for aggregation and comparison, and subscribe to message streams or publish actions back to the system. These require access to \emph{file storage, SQL and NoSQL databases, and publish-subscribe messaging systems}. Often, these may be hosted as part of the Cloud platforms themselves.  


The tasks from the above categories, along with other domain-specific tasks, are composed together to form streaming IoT dataflows. These domain dataflows themselves fall into specific classes based on common use-case scenarios, and loosely map to the Observe-Orient-Decide-Act (OODA) phases.

\emph{Extract-Transform-Load (ETL) and Archival} applications are front-line ``observation'' dataflows that receive and pre-process the data streams, and if necessary, archive a copy of the data offline. Pre-processing may perform data format transformations, normalize the units of observations, data quality checks to remove invalid data, interpolate missing data items, and temporally reorder messages arriving from different streams. The pre-processed data may be archived to table storage, and passed onto subsequent dataflow for further analysis.

\emph{Summarization and Visualization} applications perform statistical aggregation and analytics over the data streams to understand the behavior of the IoT system at a coarser granularity. Such summarization can give the high-level pulse of the system, and help ``orient'' the decision making to the current situation. These tasks are often succeeded by visualizations tasks in the dataflow to present it to end-users and decision makers.

\emph{Prediction and Pattern Detection} applications help determine the future state of the IoT system and ``decide'' if any reaction is required. They identify patterns of interest that may indicate the need for a correction, or trends based on current behavior that require preemptive actions. For e.g., a trend that indicates an unsustainably growing load on a smart power grid may decide to preemptively shed load, or a detection that the heart-rate from a fitness watch is dangerously high may trigger a slowdown in physical activities.

\emph{Classification and notification} applications determine specific ``actions'' that are required and communicate them to the IoT system. Decisions may be mapped to specific actions, and the entities in the IoT system that can enact those be notified. For e.g., the need for load shedding in the power grid may map to specific residents to request the curtailment from, or the need to reduce physical activities may lead to a treadmill being notified to reduce the speed.

\subsection{IoT Data Stream Characteristics}
IoT data streams are often generated by physical sensors that observe physical systems or the environment. As a result, they are typically time-series data that are generated periodically by the sensors. The sampling rate for these sensors may vary from once a day to hundreds per second, depending on the domain. The number of sensors themselves may vary from a few hundred to millions as well. IoT applications like smart power grids can generate high frequency plug load data at thousands of messages/sec from a small cluster of residents, or low frequency data from a large set of sensors, such as in smart transportation or environmental sensing. As a result, we may encounter a wide range of input throughputs from $10^{-2}$ to $10^{5}$ messages/sec. In comparison, streaming web applications like Twitter deal with $6000$~tweets/sec from 300M users.

At the same time, this event rate itself may not be uniform across time. Sensors may also be configured to emit data only when there is a change in observed value, rather than unnecessarily transmitting data that has not changed. This helps conserve network bandwidth and power for constrained devices when the observations are slow changing. Further, if data freshness is not critical to the application, they may sample at high rate but transmit at low rates but in a burst mode. E.g. smart meters may collecting kWh data at 15~min intervals from millions of residents but report it to the utility only a few times a day, while the FitBit smart watch syncs with the Cloud every few minutes or hours even as data is recorded every few seconds.

Message variability also comes into play when human-related activity is being tracked. Diurnal or bimodal event rates are seen with single peaks in the afternoons, or dual peaks in the morning and evening. E.g., sensors at businesses may match the former while traffic flow sensors may match the latter.

There may also be a variety of observation types from the same sensor device, or different sensor devices generating messages. These may appear in the same message as different fields, or as different data streams. This will affect both the message rate and the message size. These sensors usually send well-formed messages rather than free-text messages, using standards like SenML. Hence their sizes are likely to be deterministic if the encoding format is not considered -- text formats tend to bloat the size and also introduce size variability when mapping numbers to strings. However, social media like tweets and crowd-sourced data are occasionally used by IoT applications, and these may have more variability in message sizes.

\section{Performance  Metrics}
\label{sec:metrics}

We identify and formalize commonly-used quantitative performance measures for evaluating DSPS for the IoT workloads. 

\textbf{Latency.} 
Latency for a message that is generated by task is the time in seconds it took for that task to process one or more inputs to generate that message. If $\sigma=N:M$ is the selectivity for a task $T$, the time $\lambda^T_M$ it took to consume $N$ messages to \emph{causally produce} those $M$ output messages is the latency of the $M$ messages, with the \emph{average latency} per message given by $\overline{\lambda^T} = \frac{\lambda^T_M}{|M|}$. When we consider the average latency $\overline{\lambda}$ of the dataflow application, it is the average of the time difference between each message consumed at the source tasks and all its causally dependent messages generated at the sink tasks. 

The latency per message may vary depending on the input rate, resources allocated to the task, and the type of message being processed. 
While this task latency is the inverse of the mean throughput, the \emph{end-to-end latency} for the task within a dataflow will also include the network and queuing time to receive a tuple and transmit it downstream.

\textbf{Throughput.} 
The output throughput is the aggregated rate of output messages emitted out of the sink tasks, measured in messages per second. 
The throughput of a dataflow depends on the input throughput and the selectivity of the dataflow, provided the resource allocation and performance of the DSPS are adequate. Ideally, the output throughput $\omega^o = \sigma \times \omega^i$, where $\omega^i$ is the input throughput for a dataflow with selectivity $\sigma$. It is also useful to measure the \emph{peak throughput} that can be supported by a given application, which is the maximum stable rate that can be processed using a fixed quanta of resources. 

Both throughput and latency measurements are relevant only under \emph{stable conditions} when the DSPS can sustain a given input rate, i.e., when the latency per message and queue size on the input buffer remain constant and do not increase unsustainably.

\textbf{Jitter.} The ideal output throughput may deviate due to variable rate of the input streams, change in the paths taken by the input stream through the dataflow (e.g., at a \texttt{Hash} pattern), or performance variability of the DSPS. We use jitter to track the variation in the output throughput from the expected output throughput, defined for a time interval $t$ as, 
\[ J_t = \frac{\omega^o - \sigma \times \omega^i}{\sigma \times \overline{\omega^i}} \] where the numerator is the observed difference between the expected and actual output rate during interval $t$, and the denominator is the expected long term average output rate given a long-term average input rate $\overline{\omega^i}$. In an ideal case, jitter will tend towards zero.

\textbf{CPU and Memory Utilization.} Streaming IoT dataflows are expected to be resource intensive, and the ability of the DSPS to use the distributed resources efficiently with minimal overhead is important. This also affects the VM resources and consequent price to be paid to run the application using the given stream processing platform. 
We track the CPU and memory utilization for the dataflow as the average of the CPU and memory utilization across all the VMs that are being used by the dataflow's tasks. The per-VM information can also help identify which VMs hosting which tasks are the potential bottlenecks, and can benefit from data-parallel scale-out. 

\section{Proposed Benchmark}
\label{sec:benchmark}
We propose benchmark workloads to help evaluate the metrics discussed before for various DSPS. These benchmarks are in particular targeted for emerging IoT applications, to help them distinguish the capabilities of contemporary DSPS on Cloud computing infrastructure. The benchmarks themselves have two parts, the dataflow logic that is executed on the DSPS and the input data streams that they are executed for. We next discuss our choices for both.




\subsection{IoT Input Stream Workloads}
\begin{figure}[t]
\centering
\subfloat[CITY $@ 1000\times$ \emph{msg/sec}]{%
  \includegraphics[width=0.43\columnwidth]{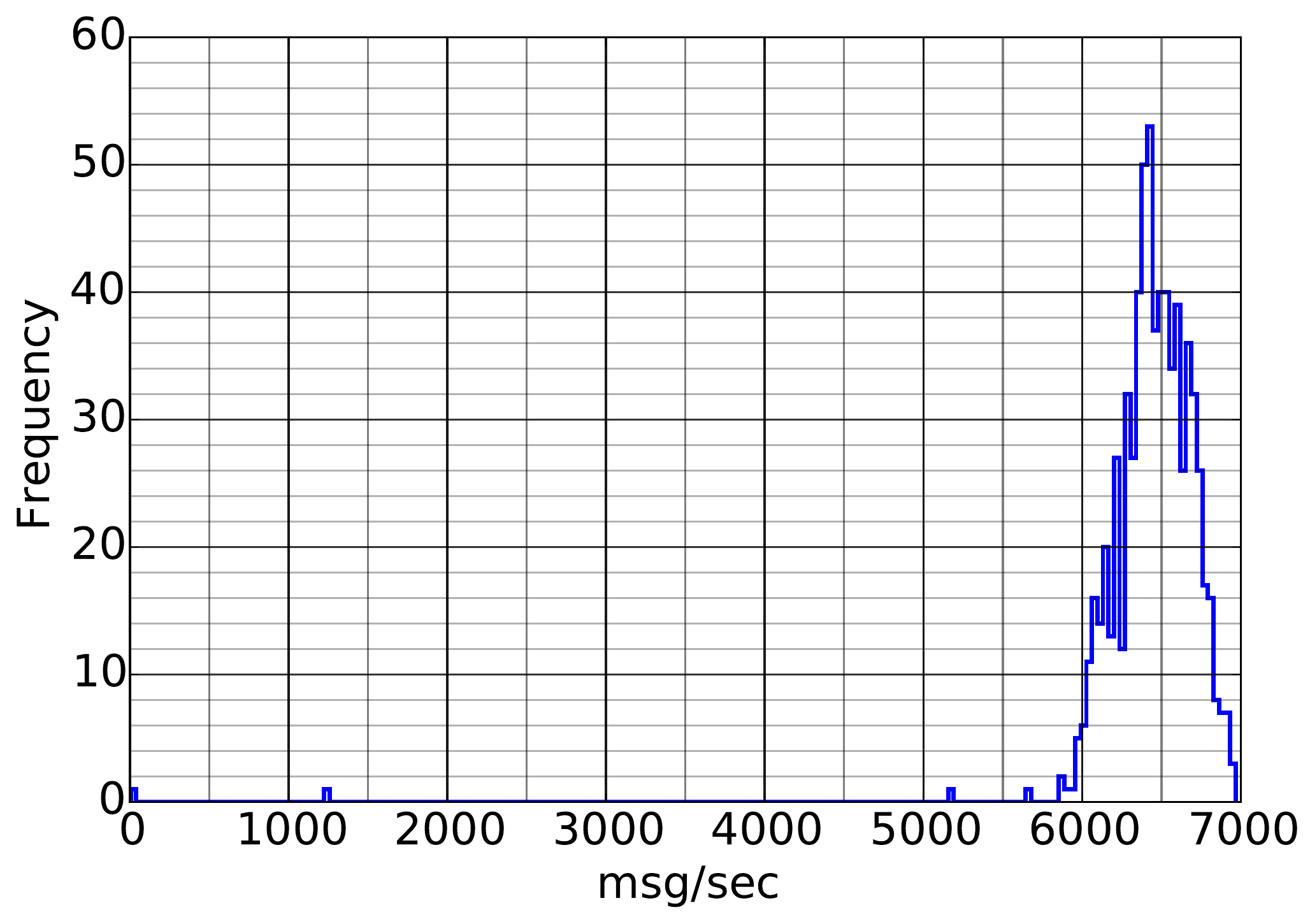}%
  \label{fig:data:sys}%
  }
  \subfloat[TAXI $@ 1000\times$ \emph{msg/sec}]{%
  \includegraphics[width=0.43\columnwidth]{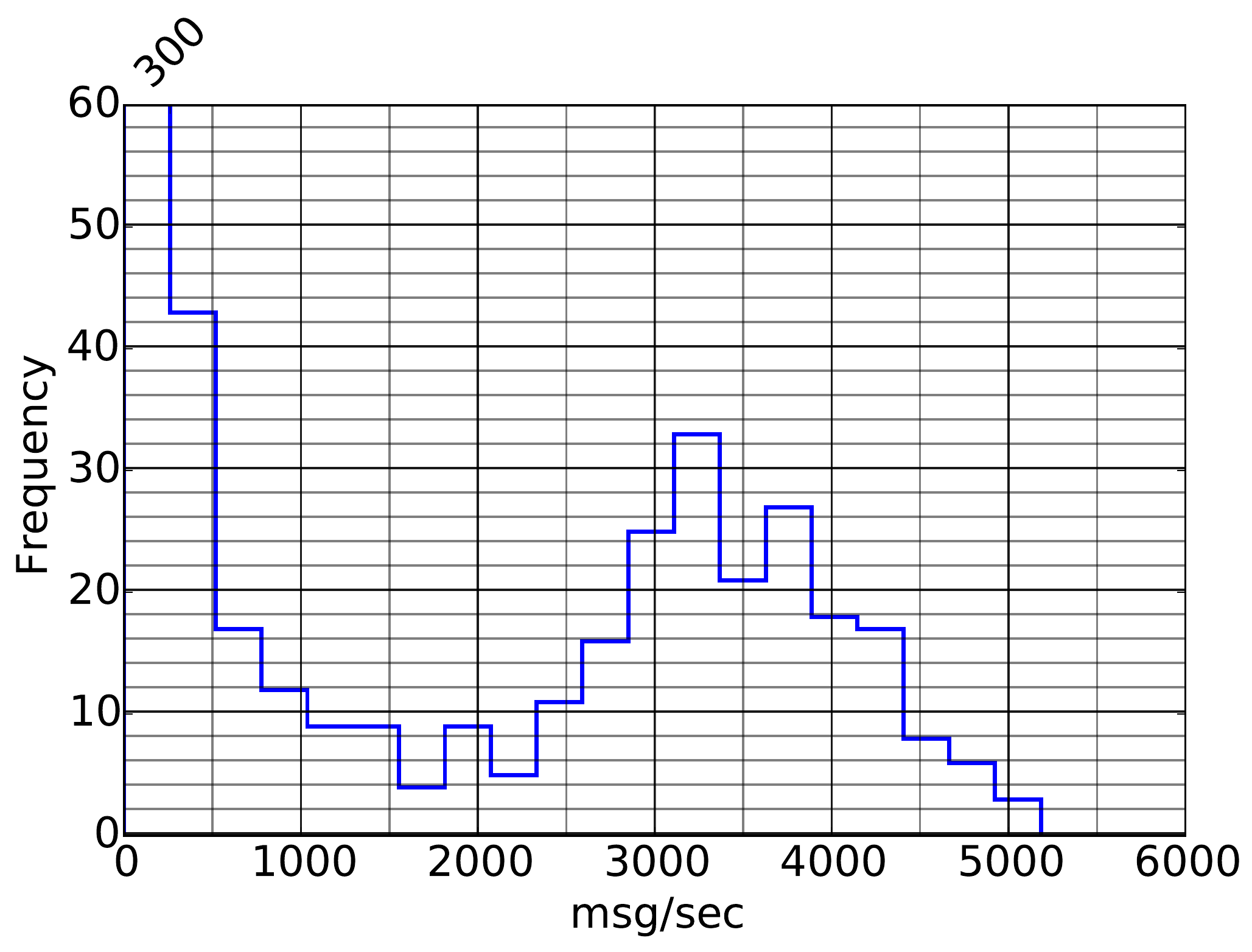}%
  \label{fig:data:taxi}%
  }
  
\caption{Frequency distribution of input throughputs for CITY and TAXI streams at $1000\times$ temporal scaling used for the benchmark runs.}
\label{fig:data-distribution}
\end{figure}

We have identified two real-world IoT data streams available in the public domain as candidates for our benchmarking workload. These correspond to smart cities domain, which a fast-growing space within IoT. Their features are shown in Table~\ref{tbl:datasets} and event rate distribution in Fig.~\ref{fig:data-distribution}.

\begin{table}[t]
\centering
\caption{Smart Cities data stream features and rates at $1000\times$ scaling}
\begin{tabular}{ccccccc}
\hline
\textbf{Dataset} & \textbf{Attributes} & \textbf{Format} & \textbf{Size}$(bytes)$ & \textbf{Peak Rate}$(msg/sec)$  & \textbf{Distribution} \\ \hline
\hline
\textbf{CITY}~\cite{data:city} & 9 & CSV & 100 & 7000 & Normal \\ \hline
\textbf{TAXI}~\cite{data:taxi} & 10 & CSV & 191 & 4000 &   Bimodal \\ \hline \hline
\end{tabular}
\label{tbl:datasets}
\end{table}

\textbf{Sense your City (CITY)~\cite{data:city}.} 
This is an \emph{urban environmental monitoring} project~\footnote{\texttt{http://map.datacanvas.org}} that has used crowd-sourcing to deploy sensors at $7$ cities across $3$ continents in 2015, with about $12$ sensors per city. Five timestamped observations, outdoor temperature, humidity, ambient light, dust and air quality, are reported every minute by each sensor along with metadata on sensor ID and geolocation. Besides urban sensing, this also captures the vagaries of using crowd-sourcing for large IoT deployments. 

We use a single logical stream that combines the global data from all $90$ sensors. Since a practical deployments of environmental sensing can easily extend to thousands of such sensors per city, we use a temporal scaling of $1000\times$ the native input rate to simulate a larger deployment of $90,000$ sensors. Fig.~\ref{fig:data:sys} shows a narrow normal distribution of the event rate centered at $6,400$~msg/sec with a peak that reaches $7,000$ msg/sec. We use 7 days of data from 27-Jan-2015 to 2-Jan-2015 for our benchmark runs.


\textbf{NYC Taxi cab (TAXI)~\cite{data:taxi}.} 
This offers a stream of \emph{smart transportation} messages that arrive from $2M$ trips taken in 2013 on $20,355$ New York city taxis equipped with GPS~\footnote{\texttt{http://www.debs2015.org/call-grand-challenge.html/}}. A message is generated when a taxi completes a single trip, and provides the taxi and license details, the start and end coordinates and timestamp, the distance traveled, and the cost, including the taxes and tolls paid. Other similar transportation datasets are also available~\footnote{\texttt{https://github.com/fivethirtyeight/uber-tlc-foil-response}}, though we chose ours based on the richness of the fields. 

Considering that events may be generated from the GPS sensors periodically rather than only at the end of the trip, we use a temporal scaling factor of $1000\times$ for our workload. This data has a bi-modal event rate distribution that reflects the morning and evening commutes, with peaks at $300$ and $3,200$~events/sec. We use 7~days of data from 14-Jan-2013 to 20-Jan-2013 for our benchmark runs. 


\subsection{IoT Micro-benchmarks}

We propose a suite of common IoT tasks that span various IoT categories, and types of streaming task patterns as well. These tasks form independent micro-benchmarks, and are further composed into application benchmarks in the next section. The goal of the micro-benchmarks is to evaluate the performance of the DSPS for individual IoT tasks, using the \emph{peak input throughput} that they can sustain on a unit computing resource as the performance measure. This offers a baseline for comparison with other DSPS, as well as when these tasks are used in application benchmarks with variable input rates.

\begin{table}[t]
\centering
\caption{IoT Micro-benchmark Tasks with different IoT Categories and DSPS Patterns}
\begin{tabular}{lccccc}
\hline
\textbf{Task Name} &~\textbf{Code}~&~\textbf{Category}~& \textbf{Pattern} &~\textbf{$\sigma$ Ratio}~&~\textbf{State}\\ \hline
\hline
XML Parsing & XML & Parse & Transform &1:1&No\\
\hline
Bloom Filter~\cite{bloom:acm:1970} & BLF & Filter & Filter &1:0/1&No\\
\hline
Average & AVG& Statistical  & Aggregate &N:1&Yes\\
Distinct Appox. Count~\cite{durand:esa:2003} & DAC & Statistical & Transform &1:1&Yes\\
Kalman Filter~\cite{kalman:asme:1959} & KAL & Statistical & Transform &1:1&Yes\\
Second Order Moment~\cite{alon:stoc:1996} & SOM & Statistical & Transform &1:1&Yes\\
\hline
Decision Tree Classify~\cite{quinlan:ml:1986} & DTC  & Predictive & Transform &1:1&No\\
Multi-variate Linear Reg. & MLR& Predictive & Transform &1:1&No\\
Sliding Linear Regression & SLR& Predictive & Flat Map &N:M&Yes\\
\hline
Azure Blob D/L & ABD& IO & Source/Transform &1:1&No\\
Azure Blob U/L & ABU& IO & Sink &1:1&No\\
Azure Table Query & ATQ& IO & Source/Transform &1:1&No\\
MQTT Publish & MQP& IO & Sink &1:1&No\\
\hline
\end{tabular}
\label{tbl:tasts}
\end{table}

We include a single XML parser as a representative parsing operation within our suite. The Bloom filter is a more practical filter operation for large discrete datasets, and we prefer that to a simple value range filter. We have several statistical analytics and aggregation tasks. These span simple averaging over a single attribute value to and second order moments over time-series values, to Kalman filter for denoising of sensor data and approximate count of distinct values for large discrete attribute values.

Predictive analytics using a multi-variate linear regression model that is trained offline and a sliding window univariate model that is trained online are included. A decision tree machine learning for discrete attribute values is also used for classification, based on offline training. Lastly, we have several IO tasks for reading and writing to Cloud file and NoSQL storage, and to publish to an MQTT publish-subscribe broker for notifications. Due to limited space, we skip pattern matching and visual analytics task categories.

A micro-benchmark dataflow is composed for each of these tasks as a sequence of a source task, the benchmark task and a sink task. As can be seen, these tasks also capture different dataflow patterns such as transform, filter, aggregate, flat map, source and sink.

\subsection{IoT Application Benchmarks}
Application benchmarks are valuable in understanding how non-trivial and meaningful IoT applications behave on DSPS. Application dataflows for a domain are most representative when they are constructed based on real or realistic application logic, rather than synthetic tasks. In case applications use highly-custom logic or proprietary libraries, this may not be feasible or reusable as a community benchmark. However, many of the common IoT tasks we have proposed earlier are naturally composable into application benchmarks that satisfy the requirements of a OODA decision making loop. 

\begin{figure}[t]
\centering
\subfloat[Pre-processing \& statistical summarization dataflow (STATS)]{%
  \includegraphics[width=0.9\columnwidth]{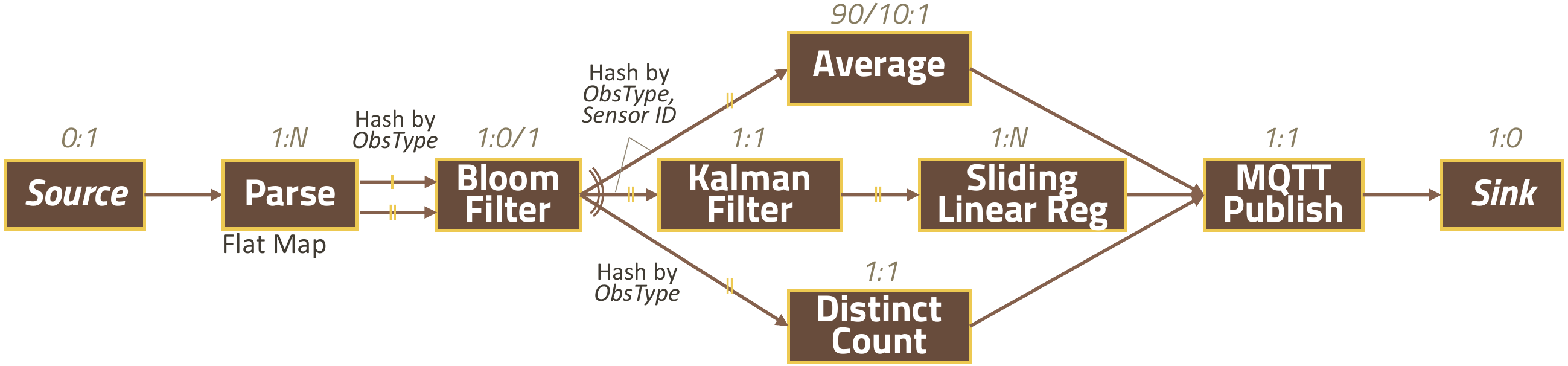}%
    \label{fig:app-stats}
  }\\
  \subfloat[Predictive Analytics dataflow (PRED)]{%
	\includegraphics[width=0.9\columnwidth]{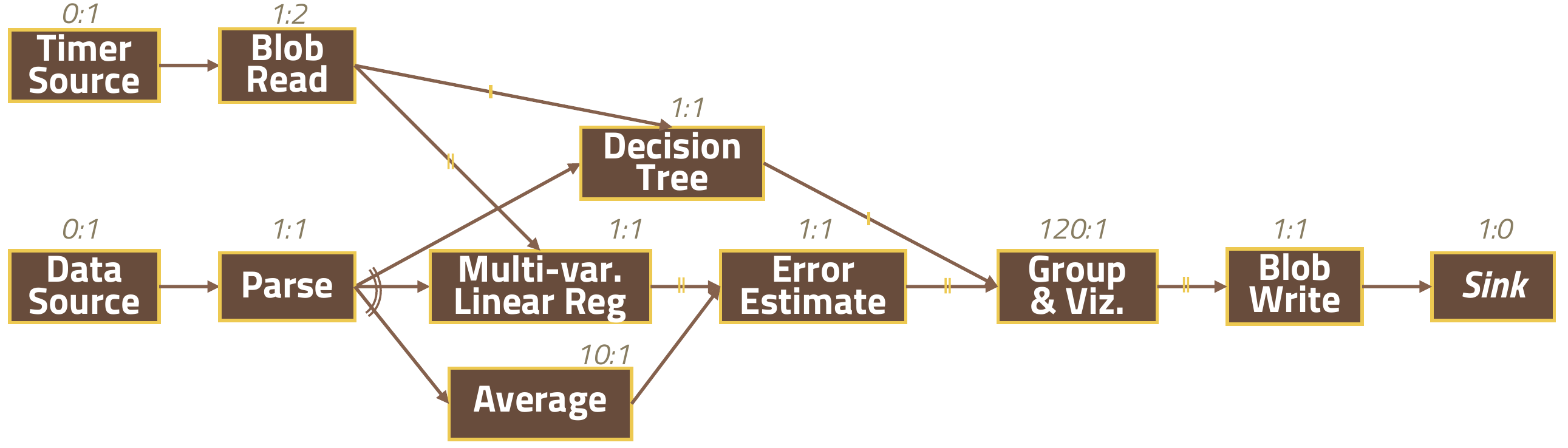}
    \label{fig:app-pred}
  }
\caption{Application benchmarks composed using the micro-benchmark tasks.}
\label{fig:app}
\end{figure}

We propose application benchmarks that capture two common IoT scenarios: a \emph{Data pre-processing and Statistical summarization (STATS)} application and a \emph{Predictive Analytics (PRED)} application. STATS (Fig.~\ref{fig:app-stats}) ingests  incoming data streams, performs data filtering of outliers on individual observation types using a Bloom filter, and then does three concurrent types of statistical analytics on observations from individual sensor/taxi IDs: sliding Average over a $90/10$ event window for CITY/TAXI ($\sim15$~mins native time window), Kalman filter for smoothing followed by a sliding window linear regression, and an approximate count of distinct readings. The outcomes from these statistics are published by an MQTT task, which can separately be subscribed to and visualized on a client browser or a mobile app. The dummy sink task is used for logging.

The PRED dataflow captures the lifecycle of online prediction and classification to drive visualization and decision making for IoT applications. It parses incoming messages and forks it to a decision tree classifier and a multi-variate regression task. The decision tree uses a trained model to classify messages into classes, such as good, average or poor air quality, based on one or more of their attribute values. The linear regression uses a trained model to predict an attribute value in the message using several others. It then estimates the error $\frac{|p-o|}{\overline{o}}$ between the predicted and observed value, normalized by the sliding average of the observations. These outputs are then grouped and plotted, and the output file written to Cloud storage for hosting on a portal. One realistic addition is the use of a separate stream to periodically download newly trained classification and regression models from Cloud storage, and push them to the prediction tasks. 


As such, these applications leverage many of the compositional capabilities of DSPS. The dataflows include \emph{single and dual sources}, tasks that are \emph{composed sequentially and in parallel}, \emph{stateful and stateless} tasks, and \emph{data parallel tasks} allowing for concurrent instances. Each message in the CITY and TAXI streams contains multiple observation fields, but several of these tasks are applicable only on univariate streams and some are meaningful only from time-series data from individual sources. Thus, the initial parse task for STATS uses a \emph{flat map} pattern ($\sigma=1:N$) to create observation-specific streams early on. These streams are further passed to task instances, grouped by their observation type and optionally their sensor ID using a \emph{hash pattern}.

\section{Evaluation of Proposed Benchmarks}
\label{sec:results}

\subsection{Benchmark Implementation}
We implement the 13 micro-benchmarks as generic Java tasks that can consume and produce objects. These tasks are building blocks that can be composed into micro-dataflows and the STATS and PRED dataflows using any DSPS that is being benchmarked. To validate our proposed benchmark, we compose these dataflows on the Apache Storm open source DSPS, popular for fast-data processing, using its Java APIs. We then run these for the two stream workloads and evaluate them based on the metrics we have defined. The benchmark is available online at \texttt{https://github.com/dream-lab/bm-iot}.

In Storm, each task logic is wrapped by a \emph{bolt} that invokes the task for each incoming tuple and emits zero or more response tuples downstream. 
The dataflow is composed as a \emph{topology} that defines the edges between the bolts, and the \emph{groupings} which determine duplicate or hash semantics. 
We have implemented a scalable data-parallel event generator that acts as a source task (\emph{spout}). It loads time-series tuples from a given CSV file and replays them as an input stream to the dataflow -- at a constant rate, at the maximum possible rate, or at intervals determined by the timestamps, optionally scaled to go faster or slower. We generate random integers as tuples at maximum rate for the micro-benchmarks, and replay the original CITY and TAXI datasets at $1000\times$ their native rates for the application benchmarks, following the earlier distribution. 
\subsection{Experimental Setup}
We use Apache Storm $1.0.0$ running on OpenJDK 1.7, and hosted on \emph{CentOS} Virtual Machines (VMs) in the Singapore data center of Microsoft Azure public cloud. For the micro-benchmarks, Storm runs the task being benchmarked on one exclusive \texttt{D1} size VM ($1$ Intel Xeon E5-2660 core at 2.2~GHz, $3.5$~GiB RAM, $50$~GiB SSD), while the supporting source and sink tasks and the master service run on a \texttt{D8} size VM ($8$ Intel Xeon E5-2660 core at 2.2~GHz cores, $28$~GiB RAM, $400$~GiB SSD). 
The larger VM for the supporting tasks and services ensures that they are not the bottleneck, and helps benchmark the peak rate supported by the micro-benchmark task on a single core VM.

For the STATS and PRED application benchmark, we use \texttt{D8}~VMs for all the tasks of the dataflow, while reserving additional \texttt{D8}~VMs to exclusively run the spout and sink tasks, and master service. We determine the number of cores and data parallelism required by each task based on the peak rate supported by the task on a single core as given by the micro-benchmarks, and the peak rate seen by that task for a given dataflow application and stream workload. In some cases that are I/O bound (e.g., MQTT, Azure storage) rather than CPU bound, we run multiple task instances on a single core.

We log the ID and timestamp for each message at the source and the sink tasks in-memory to calculate the latency, throughput and jitter metrics, after correcting for skews in timestamps across VMs. 
We also sample the CPU and memory usage on all VMs every 5~secs to plot the utilization metrics. Each experiment runs for $\sim 10$~mins of wallclock time that translates to about $7$~days of event data for the CITY and TAXI datasets at $1000\times$ scaling~\footnote{Application runtime = $\frac{7~days \times 24~hours \times 60~mins \times 60~secs}{1000\times~scaling}secs = 10.08~mins$}.

\subsection{Micro-benchmark Results}
\begin{figure}[t]
\centering
  \subfloat[Peak Throughput]{
  \includegraphics[width=0.3\textwidth]{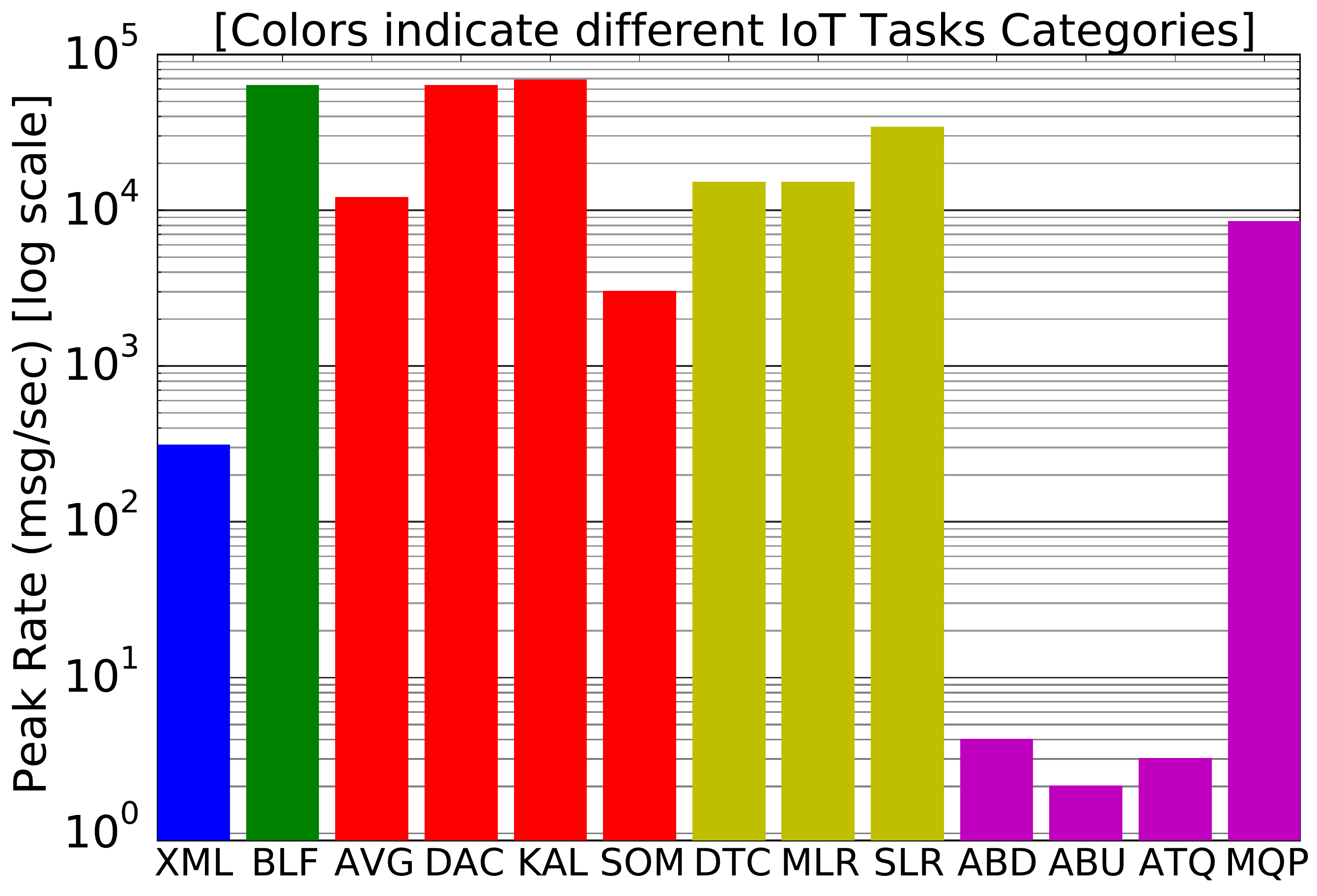}
  \label{fig:storm:micro:peakthru}
  }
  \subfloat[End-to-end latency]{
  \includegraphics[width=0.3\textwidth]{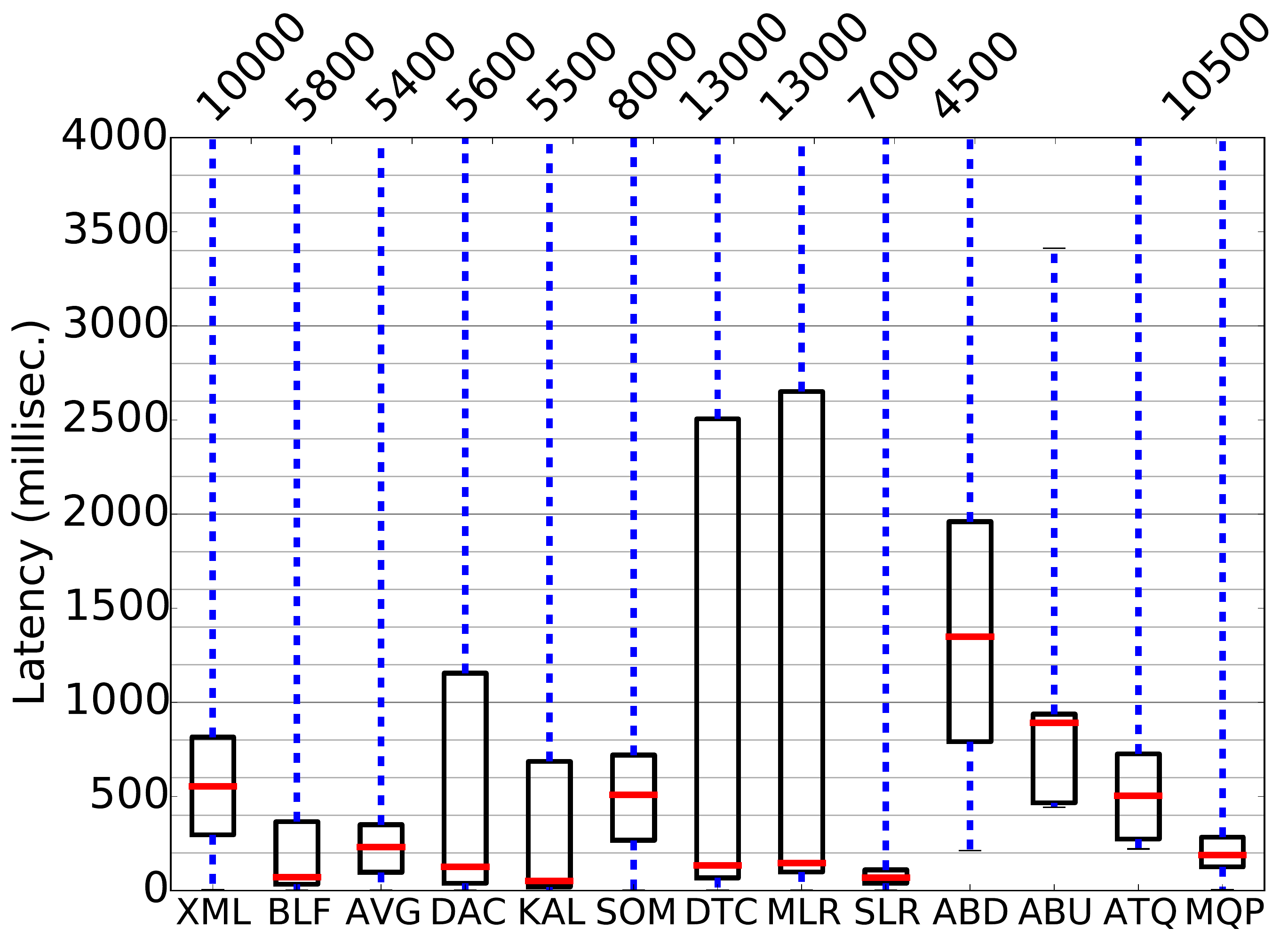}
  \label{fig:storm:micro:latency}
  }
  \subfloat[Jitter]{
  \includegraphics[width=0.3\textwidth]{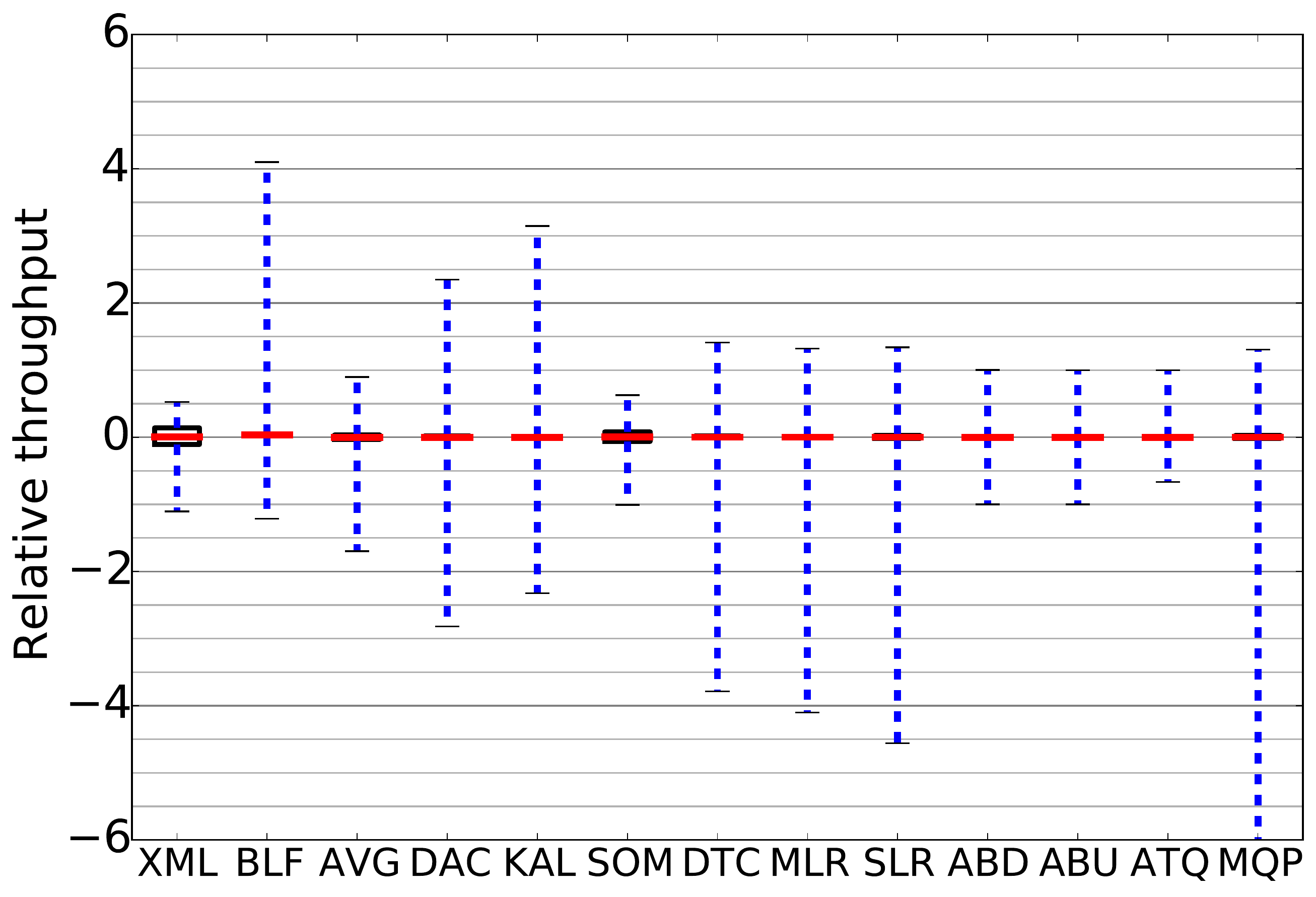}
  \label{fig:storm:micro:jitter}
  }\\
  \subfloat[CPU\%]{
  \includegraphics[width=0.30\textwidth]{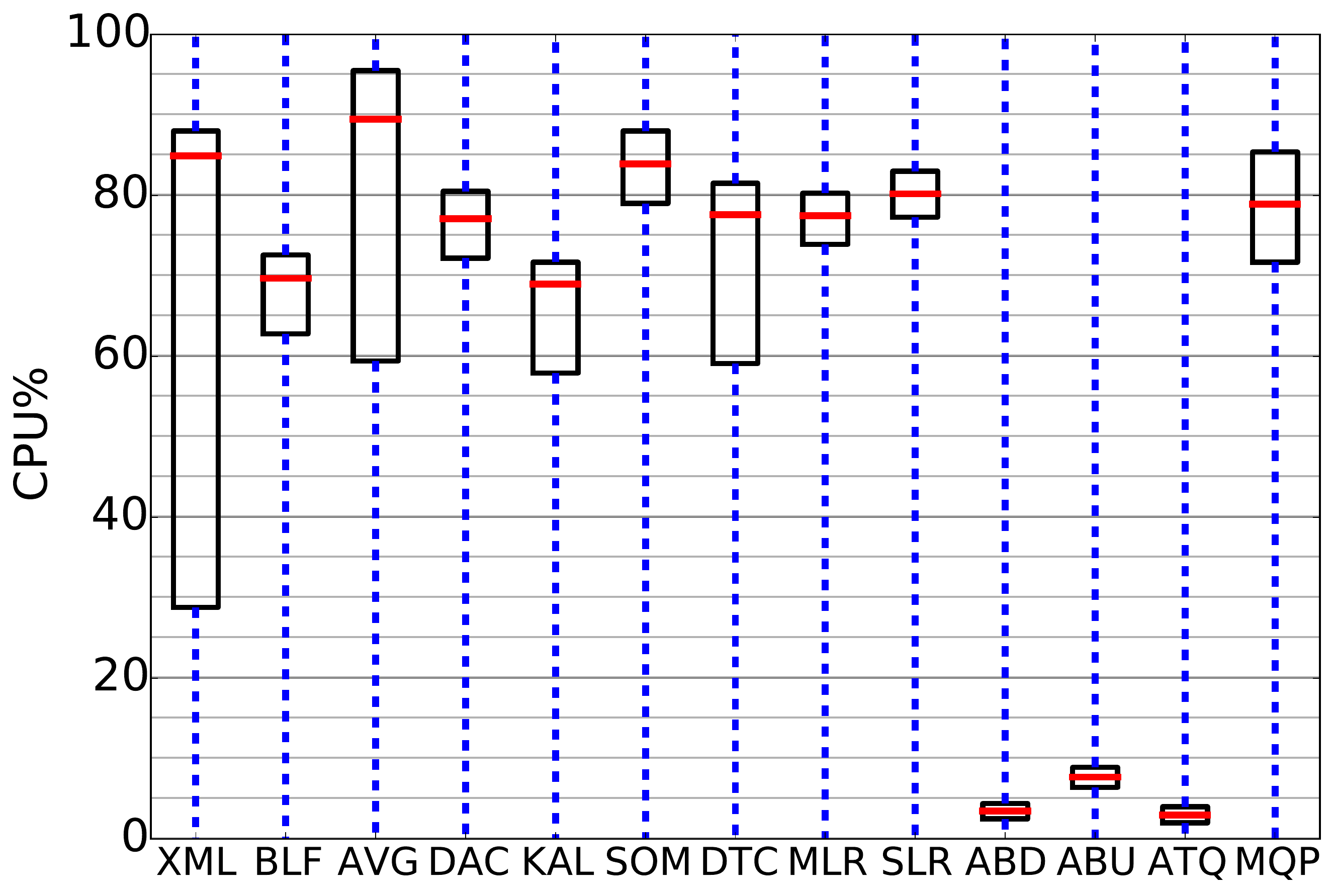}
  \label{fig:storm:micro:cpu}
  }
  \subfloat[MEM\%]{
  \includegraphics[width=0.30\textwidth]{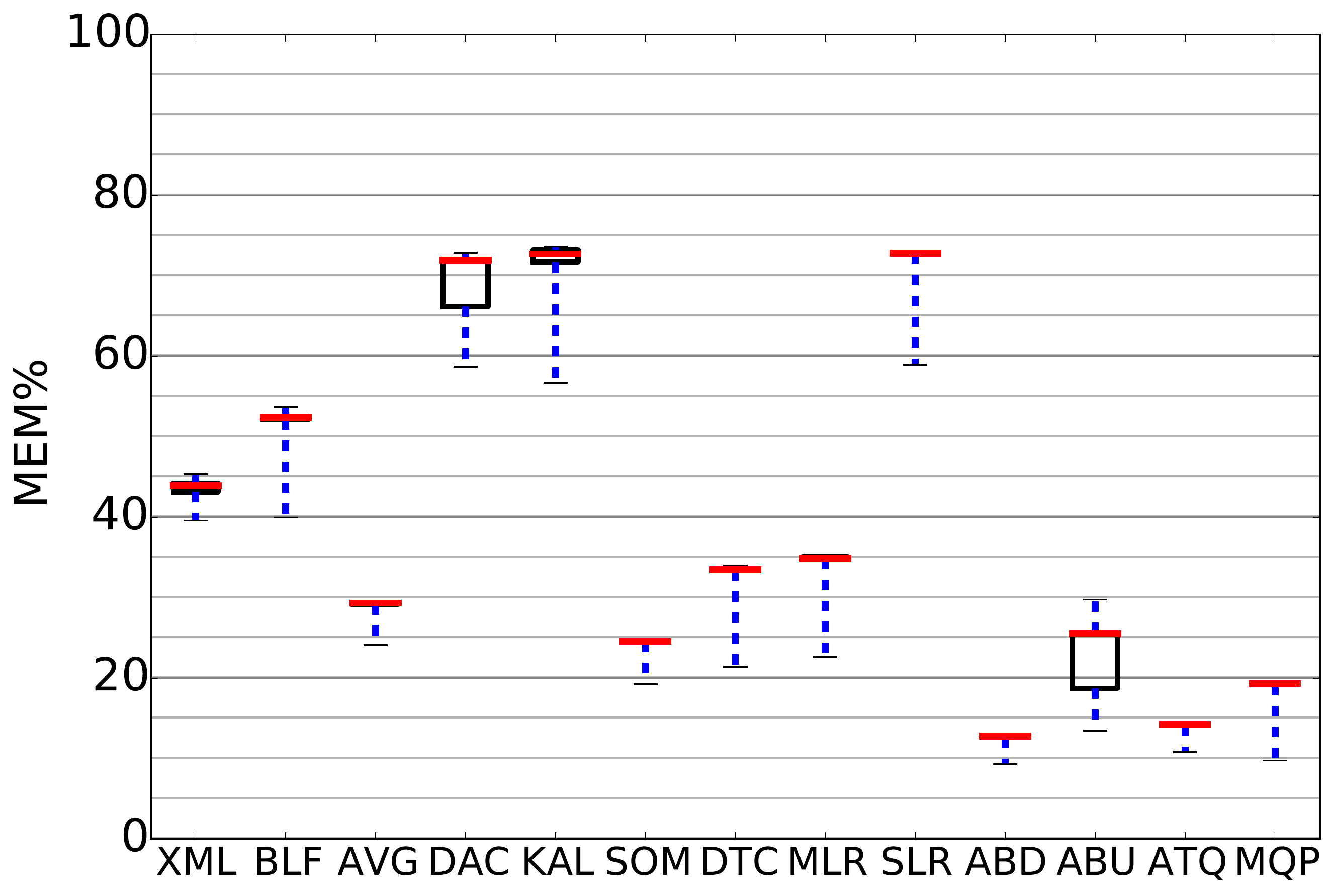}
  \label{fig:storm:micro:mem}
  }
\caption{Performance of micro-benchmark tasks for integer input stream at peak rate.} 
\label{fig:storm:micro:bm}
\end{figure}

Fig.~\ref{fig:storm:micro:bm} shows plots of the different metrics evaluated for the micro-benchmark tasks on Storm when running at their peak input rate supported on a single \texttt{D1} VM with one thread. The \emph{peak sustained throughput} per task is shown in Fig.~\ref{fig:storm:micro:peakthru} in \emph{log-scale}. We see that most tasks can support $3,000$~msg/sec or higher rate, going up to $68,000$~msg/sec for BLF, KAL and DAC. XML parsing is highly CPU bound and has a peak throughput of only $310$~msg/sec, and the Azure operations are I/O bound on the Cloud service and even slower.

The inverse of the peak sustained throughput gives the \emph{mean latency}. However, it is interesting to examine the \emph{end-to-end latency}, calculated as the time taken between emitting a message from the source, having it pass through the benchmarked task, and arrive at the sink task. This is the effective time contributed to the total tuple latency by this task running within Storm, including framework overheads. We see that while the mean latencies should be in sub-milliseconds for the observed throughputs, the box plot for end-to-end latency (Fig.~\ref{fig:storm:micro:latency}) varies widely up to $2,600$~ms for Q3. This wide variability could be because of non-uniform task execution times due to which slow executions queue up incoming tuples that suffer higher queuing time, such as for DTC and MLR that both use the WEKA library. Or tasks supporting a high input rate in the order of $10,000$~msg/sec, such as DAC and KAL, may be more sensitive to even small per-tuple overhead of the framework, say, caused by thread contention between the Storm system and worker threads, or queue synchronization. The Azure tasks that have a lower throughput also have a higher end-to-end latency, but much of which is attributable directly to the task latency.

The box-plot for \emph{jitter} (Fig.~\ref{fig:storm:micro:jitter}) shows values close to zero in all cases. This indicates the long-term stability of Storm in processing the tasks at peak rate, without unsustainable queuing of input messages. The wider whiskers indicate the occasional mismatch between the expected and observed output rates. 

The box plots for CPU utilization (Fig.~\ref{fig:storm:micro:cpu}) shows the single-core VM effectively used at $70\%$ or above in all cases except for the Azure tasks that are I/O bound. The memory utilization (Fig.~\ref{fig:storm:micro:mem}) appears to be higher for tasks that support a high throughput, potentially indicating the memory consumed by messages waiting in queue rather than consumed by the task logic itself. 


\subsection{Application Results}

\begin{figure}[t]
 \subfloat[Latency for STATS]{
 \includegraphics[width=0.235\textwidth]{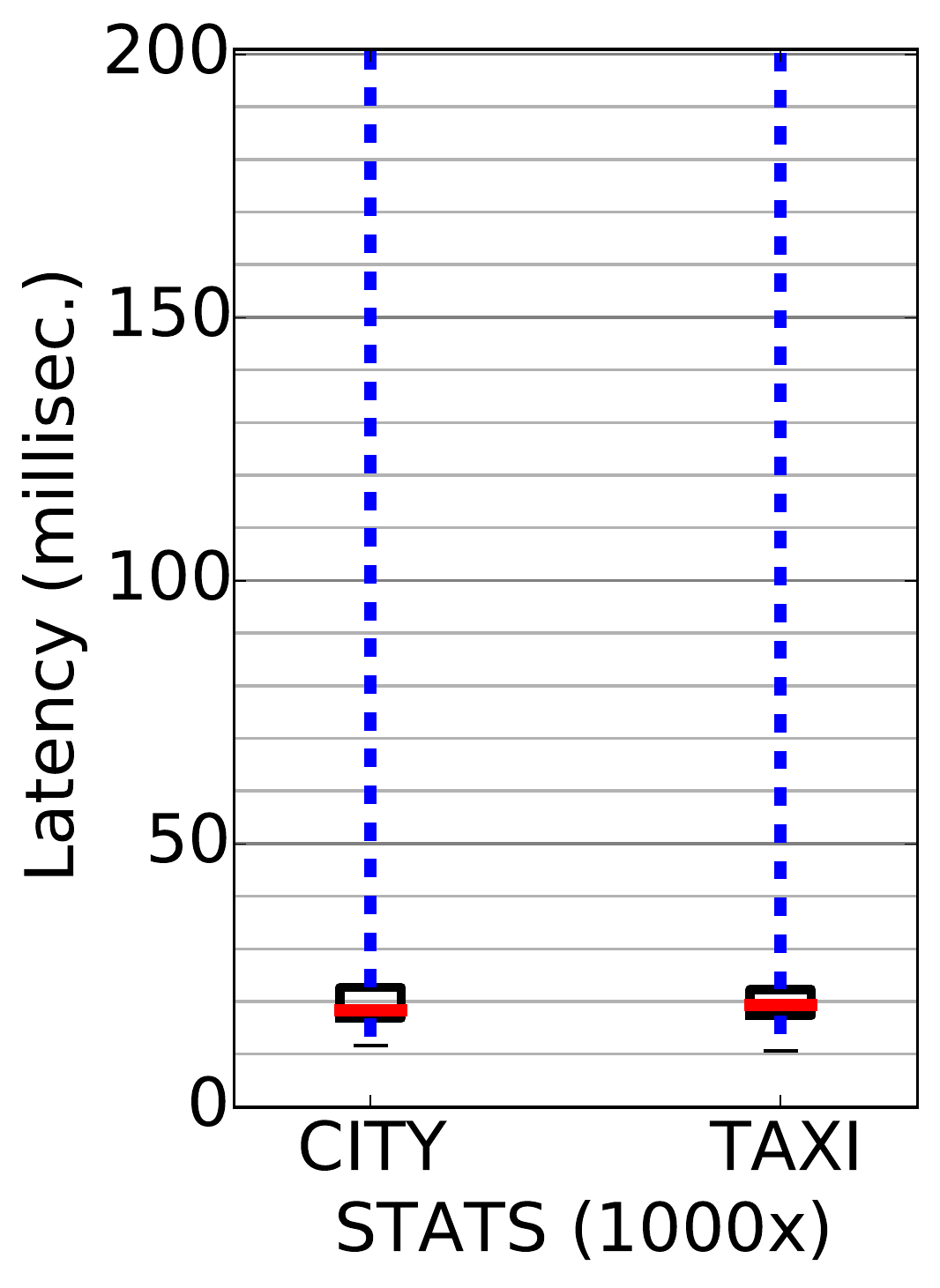}
  \label{fig:storm:stats:latency}
  }
    \subfloat[Latency for PRED]{
  \includegraphics[width=0.235\textwidth]{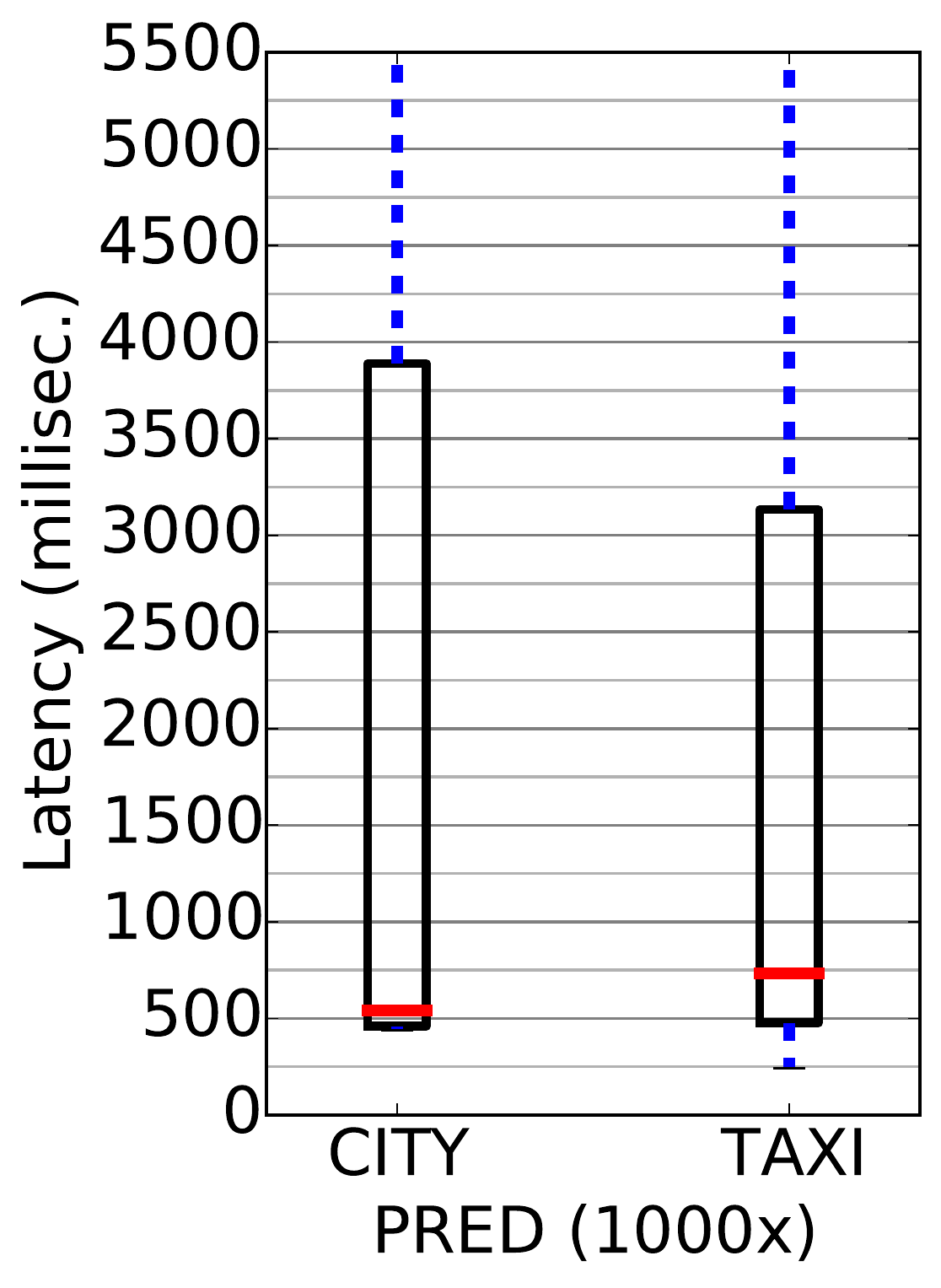}
  \label{fig:storm:pred:latency}
  }
\centering
  \subfloat[Jitter]{
  \includegraphics[width=0.46\textwidth]{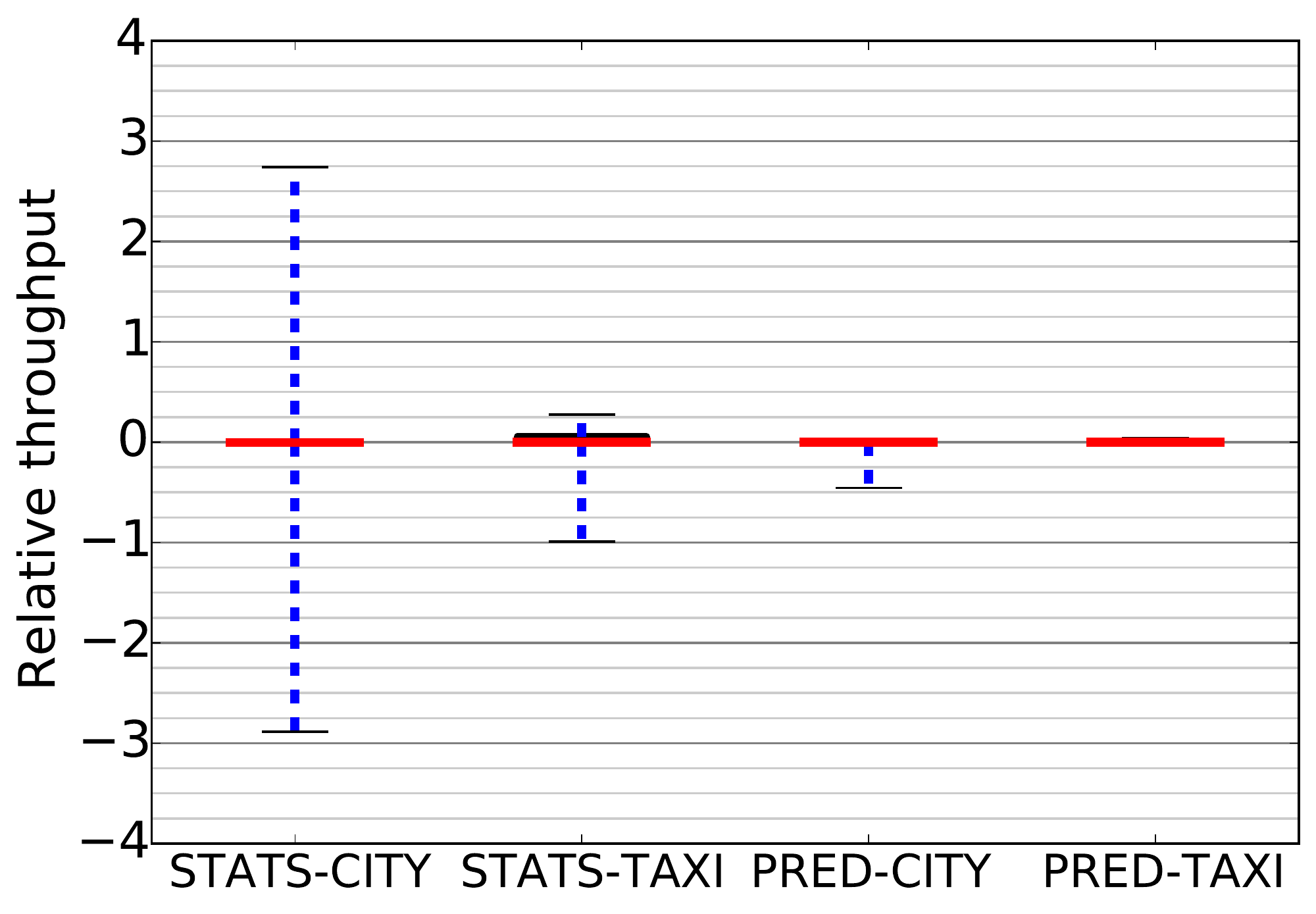}
  \label{fig:storm:app:jitter}
 } \\ 
 \subfloat[STATS:CITY]{
  \includegraphics[width=0.235\textwidth]{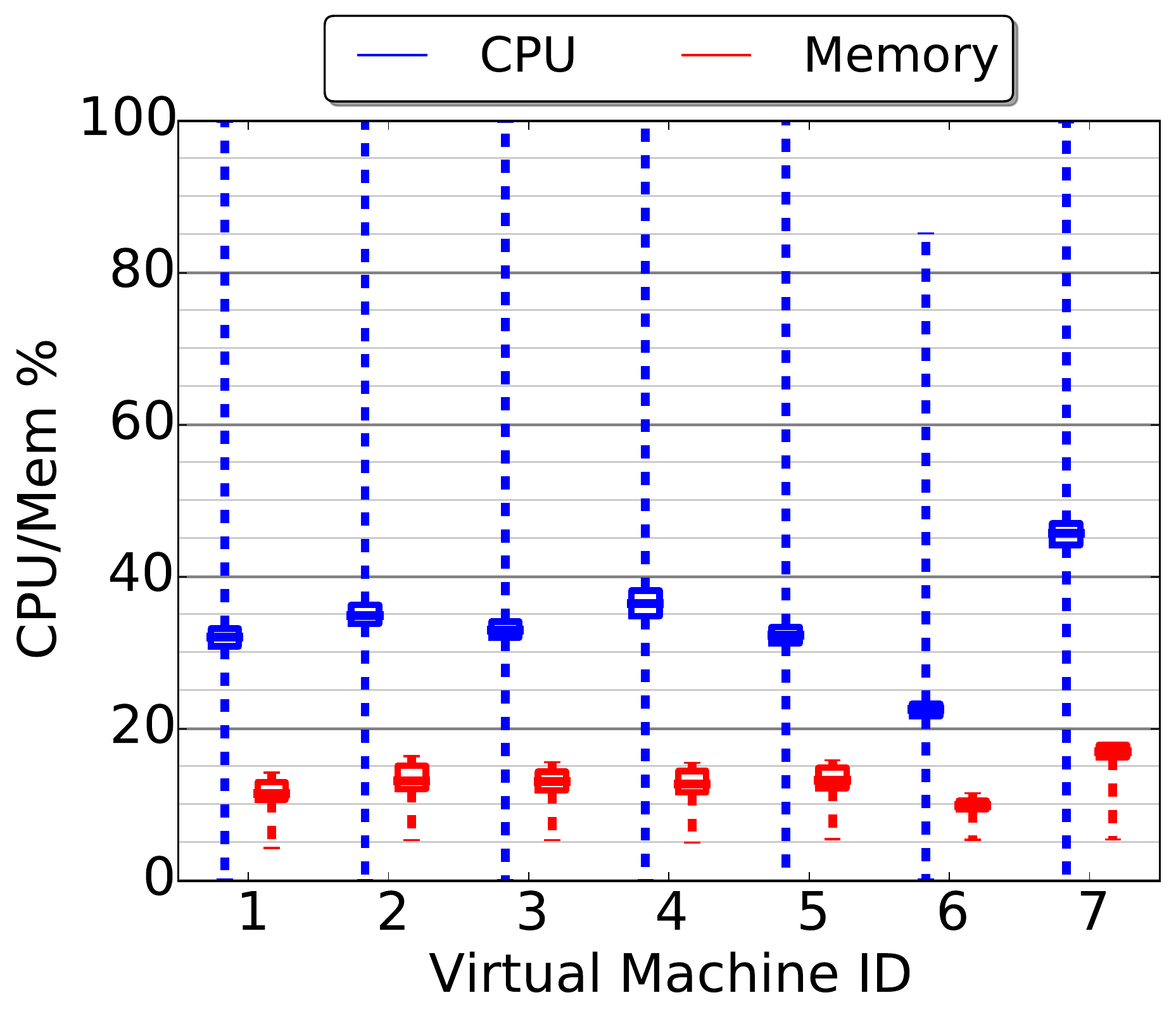}
  \label{fig:storm:stats:city:cpu}
  }
  \subfloat[STATS:TAXI]{
  \includegraphics[width=0.235\textwidth]{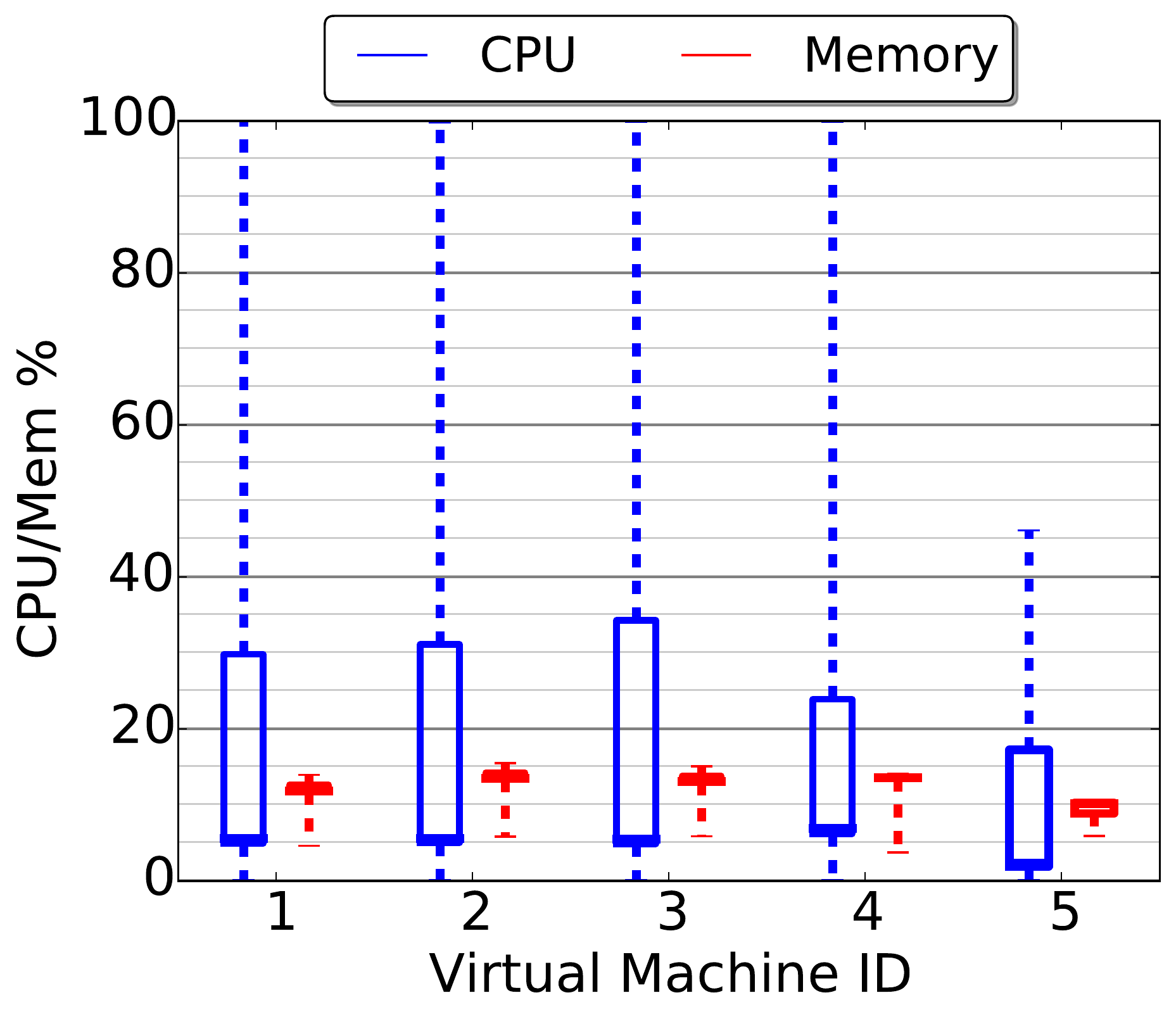}
  \label{fig:storm:stats:taxi:cpu}
  }
  \subfloat[PRED:CITY]{
  \includegraphics[width=0.235\textwidth]{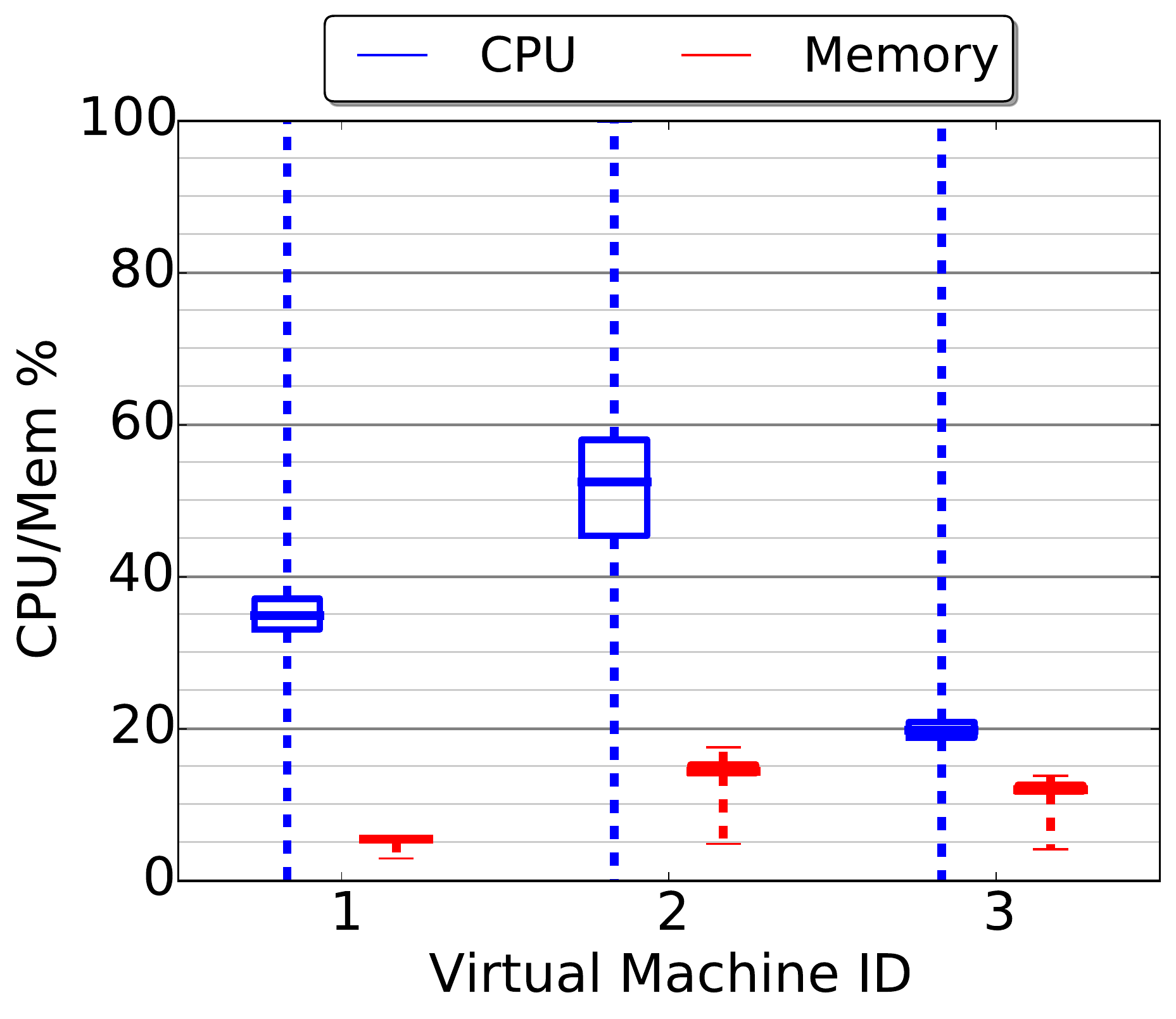}
  \label{fig:storm:pred:city:cpu}
  }
  \subfloat[PRED:TAXI]{
  \includegraphics[width=0.235\textwidth]{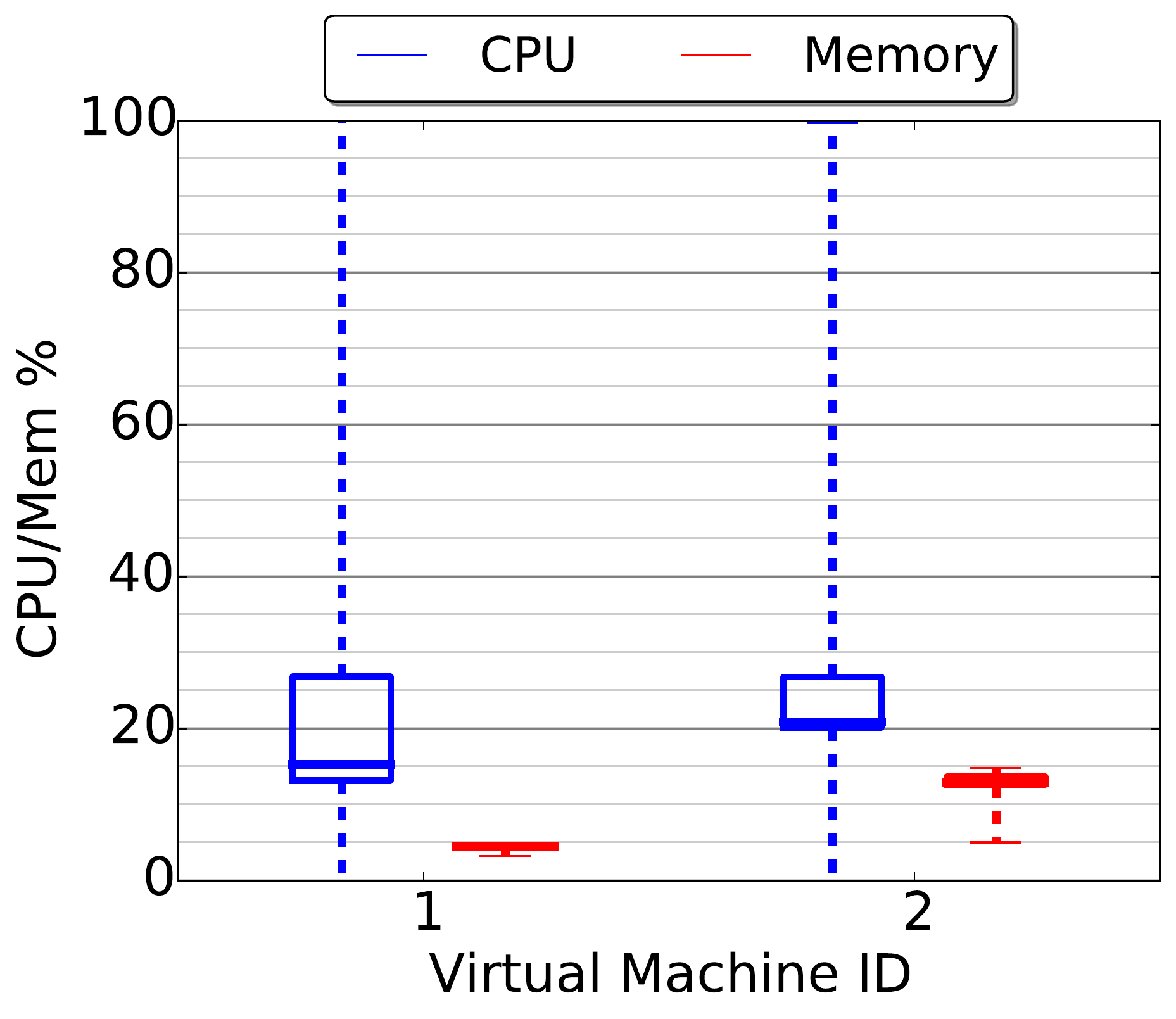}
  \label{fig:storm:pred:taxi:cpu}
  }
\caption{End-to-end latency and Jitter (top), and CPU and Memory utilization (bottom) plots for STATS and PRED application benchmarks on CITY and TAXI workloads.}
\label{fig:apps}
\end{figure}%
The STATS and PRED application benchmarks are run for the CITY and TAXI workloads at $1000\times$ their native rates, and the performance plots shown in Fig.~\ref{fig:apps}. The end-to-end latencies of the applications depend on the sum of the end-to-end latencies of each task in the critical path of the dataflow. The peak rates supported by the tasks in STATS is much higher than the input rates of CITY and TAXI. So the latency box plot for STATS is tightly bound (Fig.~\ref{fig:storm:stats:latency}) and its median much lower at $20$~ms compared to the end-to-end latency of the tasks at their peak rates. The jitter is also close to zero in all cases. So Storm can comfortably support STATS for CITY and TAXI on 7 and 5 VMs, respectively. The distribution of VM CPU utilization is also modest for STATS. CITY has a $35\%$ median with a narrow box (Fig.~\ref{fig:storm:stats:city:cpu}), while TAXI has a low $5\%$ median with a wide box (Fig.~\ref{fig:storm:stats:taxi:cpu}) -- this is due to its bi-modal distribution with low input rates at nights, with lower utilization, and high in the day with higher utilization.

For the PRED application, we see that the latency box plot is much wider, and the median end-to-end latency is between $500-700$~ms for CITY and TAXI (Fig.~\ref{fig:storm:pred:latency}). This reflects the variability in task execution times for the WEKA tasks, DTC and MLR, which was observed in the micro-benchmarks too. The Azure blob upload also adds to the absolute increase in the end-to-end time. The jitter however remains close to zero, indicating sustainable performance. The CPU utilization is also higher for PRED, reflecting the more complex task logic present in this application relative to STATS.

\section{Conclusion}
\label{sec:conclusion}
In this paper, we have proposed a novel application benchmark for evaluating distributed stream processing systems (DSPS) for the Internet of Things (IoT) domain. Fast data platforms like DSPS are integral for the rapid decision making needs of IoT applications, and our proposed workload helps evaluate their efficacy using common tasks found in IoT applications, as well as fully-functional applications for statistical summarization and predictive analytics. These are combined with two real-world data streams from smart transportation and urban monitoring domains of IoT. The proposed benchmark has been validated for the highly-popular Apache Storm DSPS, and the performance metrics presented.

As future work, we propose to add further depth to some of the IoT task categories such as parsing and analytics, and also add two further applications on archiving real-time data and detecting online patterns. We also plan to include more data stream workloads having different temporal distributions and from other IoT domains, with a possible generalization of the distributions to allow for synthetic data generation. The benchmark can also be used to evaluate other popular DSPS such as Apache Spark Streaming. 

\section*{Acknowledgments}
We acknowledge detailed inputs provided by Tarun Sharma of NVIDIA Corp. and formerly from IISc in preparing this paper. The experiments on Microsoft Azure were supported through a grant from Azure for Research.



\bibliographystyle{splncs03}
\footnotesize{
\bibliography{article}
}

\end{document}